\documentclass[journal]{IEEEtran}
\usepackage{cite}
\ifCLASSINFOpdf
   \usepackage[pdftex]{graphicx}
\else
\fi
\usepackage{amsmath}
\usepackage{algorithmic}
\usepackage[ruled]{algorithm2e} % For algorithms
\ifCLASSOPTIONcompsoc
  \usepackage[caption=false,font=normalsize,labelfont=sf,textfont=sf]{subfig}
\else
  \usepackage[caption=false,font=footnotesize]{subfig}
\fi
\usepackage{balance}
\usepackage{amssymb}
\usepackage{todonotes}
\usepackage[utf8]{inputenc}
\usepackage[english]{babel}
\usepackage{breqn}
\usepackage{textcomp}
\usepackage{hyperref}
\usepackage[inline]{enumitem}
\usepackage{color,soul}
\usepackage{makecell}
\usepackage{multirow}
\usepackage{cleveref}
\soulregister\cite7
\soulregister\ref7
\soulregister\pageref7

\begin{document}
\title{A Machine Learning-based Approach to Detect Threats in Bio-Cyber DNA Storage Systems}
%\title{Using Machine Learning to Detect Threats for Bio-Cyber DNA Storage System}
%\title{Breaking Down the Wall: Convergence of Cyber and Molecular Communication Threats in the Internet of Bio-Nano Things}

\author{Federico~Tavella, Alberto~Giaretta, Mauro~Conti,~\IEEEmembership{Senior~Member,~IEEE}, Sasitharan~Balasubramaniam,~\IEEEmembership{Senior~Member,~IEEE}
\thanks{Copyright \textcopyright 2018 IEEE. Personal use of this material is permitted.
However, permission to use this material for any other purposes must be
obtained from the IEEE by sending a request to pubs-permissions@ieee.org.}% <-this % stops a space
\thanks{F. Tavella and M. Conti are with the Department of Mathematics, University
of Padua, Padua, Italy.
E-mail: federico.tavella@studenti.unipd.it, conti@math.unipd.it}% <-this % stops a space
\thanks{A. Giaretta is with the Department of Science and Technology, Centre for
Applied Autonomous Sensor Systems, {\"O}rebro University, {\"O}rebro, Sweden.
E-mail: alberto.giaretta@oru.se}
\thanks{S. Balasubramaniam is with the Department
of Electronic and Communication Engineering, Tampere University of Technology, Tampere, Finland and Telecommunications Software \& Systems Group (TSSG), Waterford Institute of Technology, Ireland. E-mail: sasi.bala@tut.fi}}%

\maketitle

\begin{abstract}
Data storage is one of the main computing issues of this century. Not only storage devices are converging to strict physical limits, but also the amount of data generated by users is growing at an unbelievable rate. %To face these challenges, during the past decades we have witnessed a constant increase of data centres, but this explosion brings up some issues, such as the impact on the environment.
To face these challenges, data centres grew constantly over the past decades. However, this growth comes with a price, particularly from the environmental point of view.
Among various promising media, DNA is one of the most fascinating candidate. In our previous work, we have proposed an automated archival architecture which uses bioengineered bacteria to store and retrieve data, previously encoded into DNA. This storage technique is one example of how biological media can deliver power-efficient storing solutions. 
The similarities between these biological media and classical ones can also be a drawback, as malicious parties might replicate traditional attacks on the former archival system, using biological instruments and techniques. 

In this paper, first we analyse the main characteristics of our storage system and the different types of attacks that could be executed on it. Then, aiming at identifying on-going attacks, we propose and evaluate detection techniques, which rely on traditional metrics and machine learning algorithms. We identify and adapt two suitable metrics for this purpose, namely \textit{generalized entropy} and \textit{information distance}. Moreover, our trained models achieve an AUROC over 0.99 and AUPRC over 0.91.
\end{abstract}


% \fi
\IEEEpeerreviewmaketitle

\section{Introduction}

\IEEEPARstart{T}{he} World Wide Web~\cite{www} has transformed the way human beings create and share information, breaking down cultural barriers. Emerging communication technologies (e.g., 5G and Internet of Things) enabled seamless connectivity and novel applications. For example, social networks allow people to share their experience through messages, pictures, audio, and video recordings. As a consequence, we face an increasingly amount of data, bound to further grow as more and more devices will connect to the Internet in the future.
Currently all these generated data are stored in large data centres, %allowing different kind of users (e.g., common people, professionists or corporations) to store all their data without the cost of a physical device. 
resulting in large investments in cloud services and infrastructures~\cite{Facebook}. %does not mean that these physical devices do not need to be produced: the number of 
%as well as data centers, which has been major growth in recent years. 
%While future expansion of these infrastructures is necessary and inevitable, they will also brings along a number of challenges. Data centres are known to consume phenomenal amounts of energy, which takes a relevant toll from the environmental impact point of view~\cite{EnvironmentTime}~\cite{EnvironmentNYTimes}. Besides, the energy requirements for powering as well as cooling, the data centres also place immense strains on the operation costs. Driven by these challenges, researchers have been exploring alternative storage mediums.

While these infrastructures are necessary and inevitable, they also bring along some challenges. Data centres consume phenomenal amounts of energy and heavily impact the environment~\cite{EnvironmentTime}~\cite{EnvironmentNYTimes}. Besides, the energy required for powering and cooling the data centres place immense strain on the operation costs. Driven by these challenges, researchers have been exploring alternative storage mediums.

%An emerging and promising approach is storing data into \textbf{DNA}. From a computing perspective, DNA represents a biological cell's software and determines the functionalities that maintains stability within an organism. This software instruction set, contains vast amount of information that enables different types of cells to operate and function as a collective system for different tissues as well as organs. This key characteristic can allow us to exploit its use to enable data storage, where information can be encoded into a series of nucleotides and represented as genes that can be inserted into DNA. There have been a number of proposed techniques for encoding information into genetic sequences. For example, Goldman et al ~\cite{Goldman} developed a simple 2-bit encoding scheme that was able to store a large quantity of data into DNA strands. However, a major question is the future practicality of DNA storage, and whether they can be integrated into conventional data centres, and what are some of the security implications, since we are dealing with biological substrates, which in itself brings along a number of challenges. In particular, how will security be handled for a system that requires random access towards the DNA storage medium to retrieve data. %Someone may wonder how these data can be accessed, particularly in parallel, and how it is possibile to achieve the equivalent of random access in electronic storage. 

An emerging and promising approach is to store data into \textbf{Deoxyribonucleic Acid (DNA)}. From a computing perspective, DNA represents biological cells' software and determines organisms' functionalities. This software enables different types of cells to operate as a collective system, such as tissues and organs. This key characteristic allows us to use cells as data storage units, where information is encoded into nucleotides and inserted in the DNA.
There have been many proposed techniques for encoding information into genetic sequences. For example, Goldman et al.~\cite{Goldman} developed a simple 2-bit encoding scheme, capable of storing a large quantity of data into DNA strands. A major question is the future practicality of DNA storage, and how we could integrate it in conventional data centres. Since we are dealing with biological substrates, we have to be aware of attacks that use either organisms or chemicals. The complication increases with biological systems equipped with random access functionalities.
%need to handle security that deals with biological interactions for random access for DNA storage system.
An example for random access was proposed by researchers from University of Washington and Microsoft~\cite{Microsoft}. In their work, the authors proposed multiple pools of DNA storage, where each DNA strand corresponds to binary data. To support random access capabilities, every strand has a primer (i.e., a specific sequence of nucleotides used as a starting point for DNA synthesis). Tavella et al.~\cite{Tavella} proposed to store information within bioengineered bacteria~\cite{Moore11}~\cite{Moore13}~\cite{Okaie}, and retrieve it through a bacterial nanonetwork. Bacterial nanonetworks are artificial networks that exhibit molecular communication characteristics, and enable multi-hop links between motile and non-motile bacteria. 

In this paper, we focus on the security implications of Tavella et al.~\cite{Tavella} approach, where digital information is encoded into DNA plasmids, which are physically stored inside bacteria. In particular, bacteria are known to interact and communicate as part of their social ecosystem, both intraspecies and interspecies. Communication is critical for bacteria to survive collectively as a population and evolve through different environmental changes. When information is stored in bacteria, population interference can occur from other bacteria living within the same environment. Such interference need to be monitored in order to ensure that the retrieval process maintains an acceptable level of reliability. Giaretta el al. ~\cite{Giaretta} presented how security can affect bacterial nanonetworks, by blocking or deviating the movement of motile bacteria towards an intended location. %how an attacker can manipulate the environment surrounding the bioengineered bacteria to modify their behaviour.

\begin{figure}[ht]
\centering
\includegraphics[width=0.48\textwidth]{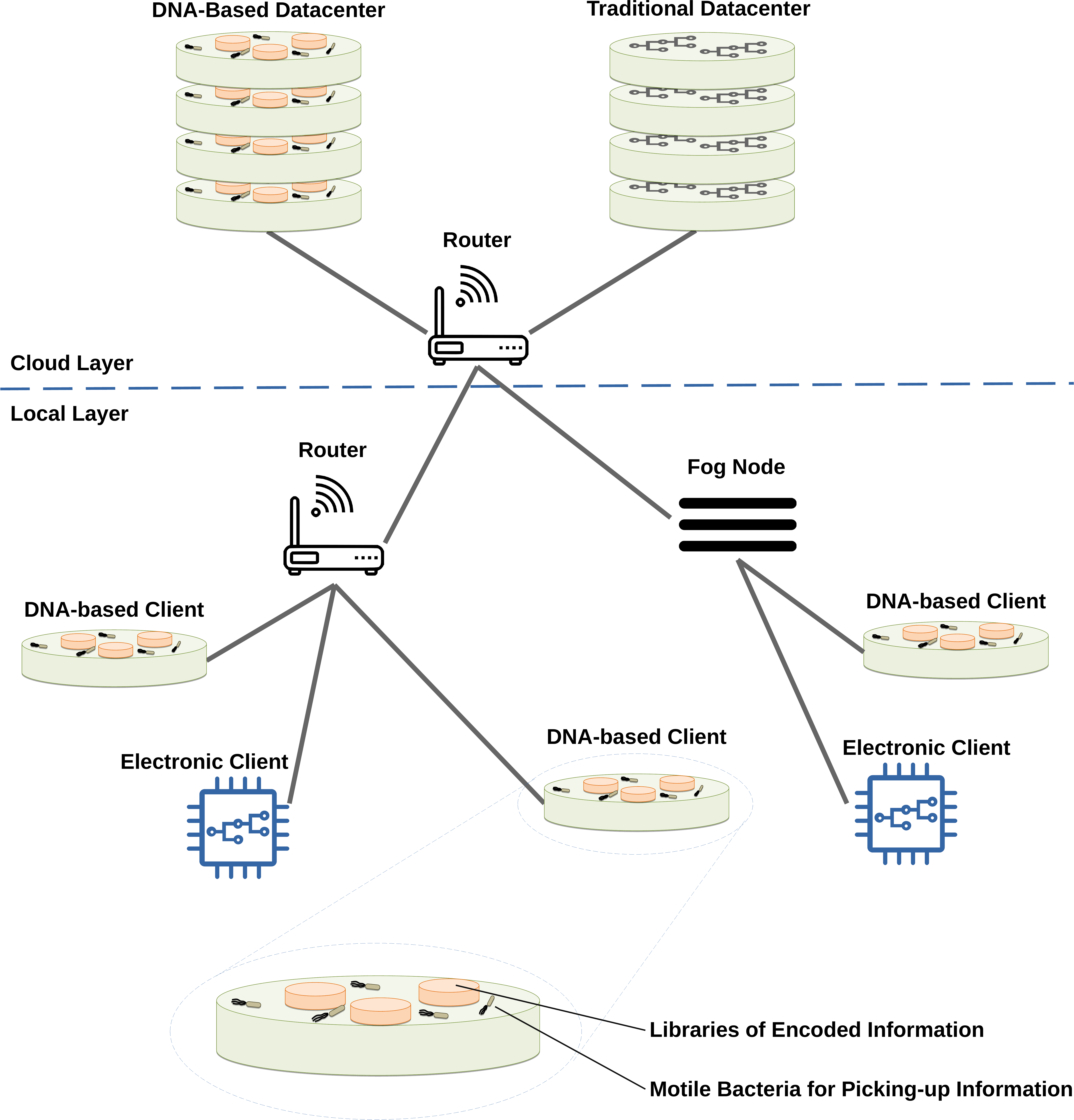}
\label{fig:network_arch}
\caption{Architecture of a mixed network, where traditional electronic devices and servers co-exist with DNA-based devices.}
\end{figure}

In Figure~\ref{fig:network_arch}, we present an architecture where bacterial nanonetworks can be integrated into conventional data centres. The bacterial storage can be located in data centres, or can be part of a remote device owned by the end user, such as an Internet of Things (IoT) device. Research showed that current IoT devices are susceptible to security attacks~\cite{HackableThings}, making them unfit candidates for holding complex entities like bacterial networks. Moreover, IoT devices generally lack in computational power, increasing the challenge in deploying monitoring algorithms to detect security attacks. Data centres are better candidates for integrating a biological infrastructure. First, they exhibit isolation which allows them to safely host a DNA storage system based on bacterial nanonetworks. Second, they have access to enough computational power for storing and running monitoring algorithms.
%On the other hand, data centres exhibit both isolation and computational power, making them the best candidates to integrate a biological infrastructure that holds bacterial nanonetworks for DNA storage system, where the silicon technology in the data centre will be used to house the monitoring algorithms.

In our study, we investigate two types of algorithm for detecting attacks on the bacterial nanonetwork DNA storage. We use information metrics, previously used for detecting Distributed Denial of Service (DDoS) attacks in conventional networks~\cite{Xiang}, and we use Machine Learning (ML) techniques.

The contributions of this paper are manifold:
\begin{enumerate}
\item An analysis of bacterial nanonetwork DNA storage system from a computer science perspective, highlighting similarities and differences with conventional storages;
\item A security attack that uses competing bacteria to disrupt the bacterial nanonetwork infrastructure;
\item Information metric techniques reformuled for detecting DDoS attacks in conventional networks, together with an assessment of such techniques for bacterial nanonetwork storages; %analyze and reformulate some metrics used in computer and network security to fit our scenario;
\item Machine learning algorithms for monitoring bacterial nanonetworks and detecting attacks; %Given the biological vectors of attack, we delineate the possible vulnerabilities of the DNA archival;
\item A thorough evaluation of all the presented techniques through extensive simulations.
%\item Finally, we define a two different detection techniques, based on \textbf{metrics} and \textbf{machine learning}, to overcome the system weaknesses and raise a warning when the system is under attack.
\end{enumerate}

We organize this paper as follows: Section~\ref{sec:rel_works} describes the DNA automated archive and the different attacks that a malicious user could execute into it. Section~\ref{sec:sec_metr} describes traditional metrics for detecting DDoS attacks and how we can adapt them to our scenario. Section~\ref{sec:ml} explains how machine learning can be used for the same purpose. In Section~\ref{sec:sim} we assess the attacks dangerousness, as well as the effectiveness of our detection techniques. Last, in Section~\ref{sec:disc} we discuss our results and draw our conclusions.

\section{System model}\label{sec:rel_works}
In this section, we describe the functioning of bacteria-based DNA-based storage devices and the corresponding attacks that can be lead onto these systems. 
%Giving a snippet of what follows in the next section, we anticipate that the usage of a storage system based on biological components can open the way to new scenarios where malicious people do not use digital tools to perform an attack on digital systems. Instead, they can use biological instruments (e.g., bacteria) to compromise digital information.

\subsection{DNA storage}\label{sec:dna_stor}
\begin{figure}[ht]
\center
\includegraphics[trim=0mm 0mm 0mm 0mm,clip,width=0.45\textwidth]{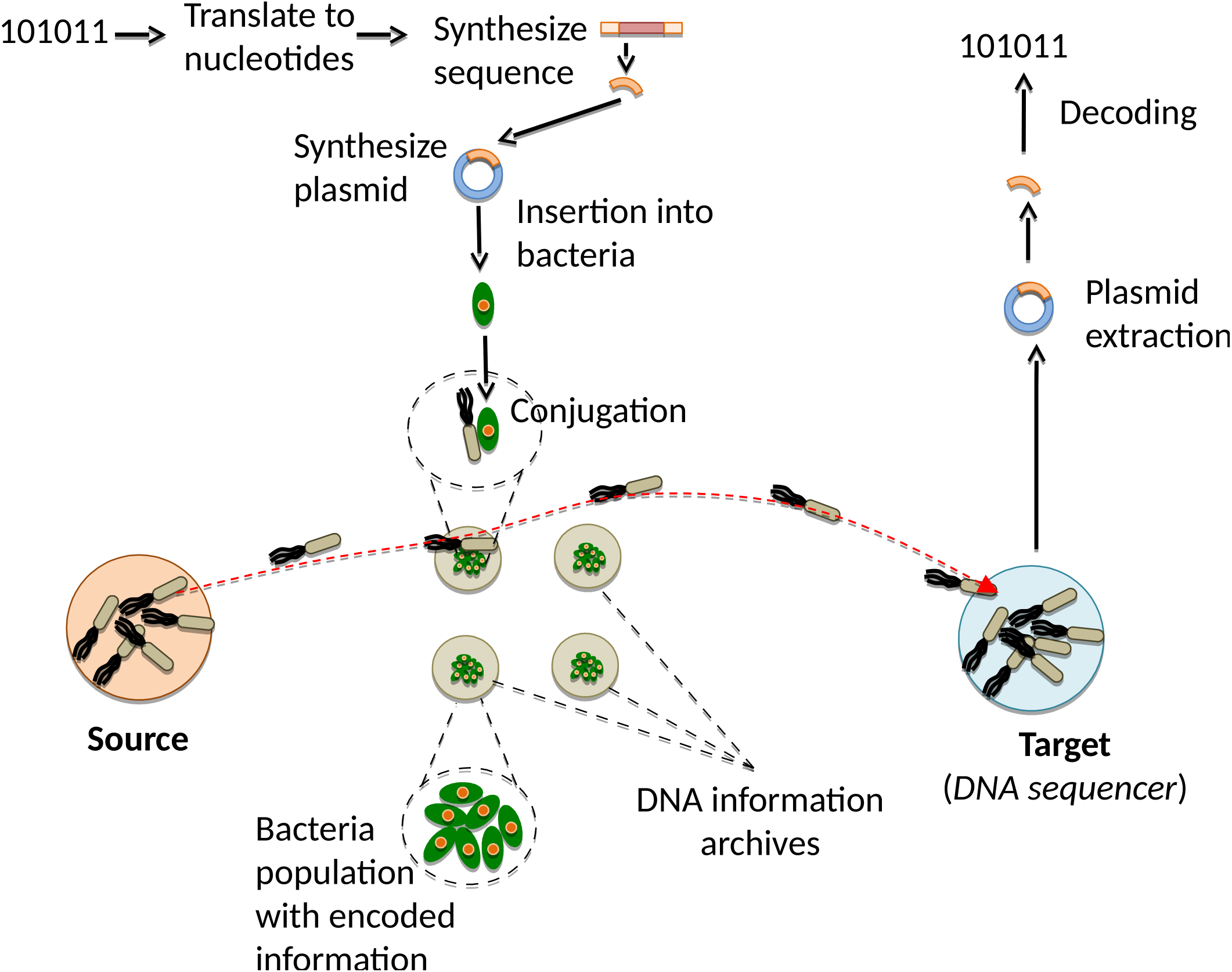}
\caption[Overall system architecture of the DNA archival system.]{Overall system architecture of the DNA archival system that enables reading using bacterial nanonetworks. Motile bacteria (in grey) swim from the source towards each of the library points, which contain different information. Reached the motile-restricted bacteria, conjugation starts to retrieve the encoded information contained into the plasmid. Last step, motile bacteria swim towards the Target, where the plasmid is retrieved through sequencing processes.}
\label{fig:System_architecture}
\end{figure}

The DNA automated archive~\cite{Tavella} is a biological storage system that combines two different components: DNA encoding and bioengineered bacteria. The first component provides a way for transferring and translating digital information into DNA, while the bacteria are used as data storage and access mechanism.
Figure~\ref{fig:System_architecture} describes the overall system architecture of the system proposed in~\cite{Tavella}. Digital information is firstly encoded into nucleotides, and for this the authors use different encoding techniques. %The synthesized genes of the encoded nucleotides are then inserted into a plasmid, and this is taken up by the bacteria through the process of transformation. Plasmids are then placed into motile-restricted bacteria, and to guarantee that they do not get mobilized, they are placed on solid agar (i.e., a jelly-like substance used to measure microorganisms' ability to move). The motility-restricted bacteria with different data are placed into specific regions of the grid as illustrated in Figure~\ref{fig:System_architecture}.
The synthesized genes of the encoded nucleotides are inserted into plasmids, and plasmids are inserted into bacteria through the transformation process. To ensure motility restriction, these bacteria are place on solid agar (i.e., a jelly-like substance used to measure microorganisms' ability to move). The motility-restricted bacteria with different data are placed into specific regions of the grid, as illustrated in Figure~\ref{fig:System_architecture}. 
In the event that a read operation needs to be performed, motile bacteria are released from \emph{A}, swim towards the compartment, and then \textit{conjugate} with the motile-restricted bacteria to retrieve the plasmids with the encoded information. Once this is complete, bacteria swim towards position \emph{C} to deliver the plasmids. Conjugation is a process where bacteria come together and form a physical connection that allows them to transfer plasmids between each other, and this process has a probability associated with it. At position \emph{C}, the plasmids are retrieved, sequenced to obtain the data, and decoded back into digital format.

A key requirement for the motile bacteria is the capability of swimming towards an accurate point. This allows to conjugate with the correct batch of motile-restricted bacteria, and retrieve the right plasmid with the encoded information. This is where the proposed Molecular Positioning System (MPS)~\cite{Tavella} plays a role. As suggested by Okaie et al.~\cite{Okaie}, it is possible to deploy chemoattractants and redirect engineered-bacteria by means of chemotaxis (i.e., movement as a response to chemical stimuli). This pairs well with MPS~\cite{Tavella}, which is based on the receptor saturation addressing technique, proposed by Moore and Nakano~\cite{Moore11}~\cite{Moore13}.

\subsection{Vulnerabilities}\label{sec:vuln}

Regardless of the fact that it is based on bacterial nanonetworks, the DNA archive~\cite{Tavella} is, to many extents, similar to electronic storage systems. Therefore, it is prone to most of the common databases attacks. Here, we propose two examples of these attacks: Denial of Service (DoS) and sniffing.
%In the following sections, we see how this is possible using two different biological ``tools'': bacteria and phages.

\subsubsection{Denial of Service}
\begin{figure}[htb]
\centering
\includegraphics[trim=0mm 0mm 0mm 0mm,clip,width=0.45\textwidth]{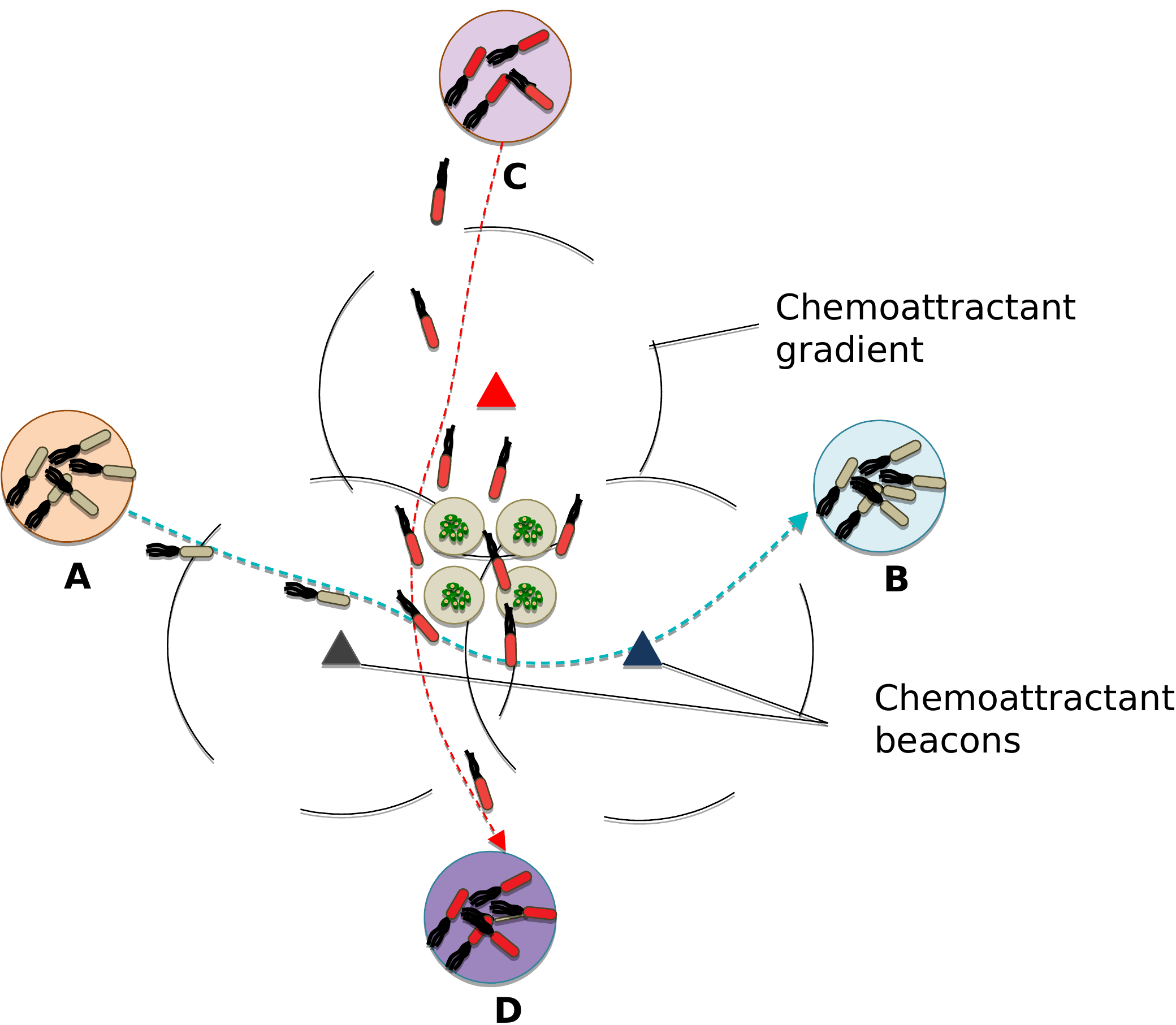}
\caption{Example of DoS scenario. Legitimate bacteria move from starting point \textit{A} to the agars containing the storage bacteria. Once the conjugation is complete, they move towards their destination \textit{B}. Malicious bacteria move from \textit{C} to \textit{D} trying to conjugate with the ones in the agar, blocking the retrival process of the legitimate bacteria.}
\label{fig:dos}
\end{figure}

In order to share DNA, bacteria use a method called \textit{conjugation}. During this process, two bacteria physically connect to each other to share plasmids (i.e., circular DNA strands). However, each bacterium can conjugate only with one bacterium. Thus, if attackers spread their own bacteria around the clusters area, the bacteria conjugate with the set of bacteria that contain the encoded data, as Figure~\ref{fig:dos} illustrates. Malicious bacteria move from point $C$ to the cluster, cluttering the system, and after completing the conjugation process reach point $D$. In this way, the legitimate bacteria that are meant to retrieve the DNA are not able to conjugate and access the information in the archive. This equals to a DoS attack on a database, since it prevents the main feature of the archive, retrieving stored data. Besides blocking and cluttering the clusters of motile bacteria, the cluttering can also occur at the destination where the bacteria are collected before they are pulled into the sequencer.

%\subsection{Re-routing}
%
%The Molecular Positioning System (MPS) described in~\cite{Tavella} uses three different beacons to build up a trilateration system. It is quite clear that a modification to the MPS scheme could affect the coordinate system used by bacteria. In order to do so, an attacker could place another beacon to make the bacteria believe that the point they should reach is in another place, which could lead to stealing the encoded information or slightly deviating bacteria and consequently slowing down the retrieving process, as illustrated in Figure~\ref{fig:spoofing}. In this way, a malicious user can replicate what happens when we change the IP address to redirect the traffic into a network. Thus, we can trace back this scenario to a spoofing attack on a digital system. The same technique can also be used to implement a DoS attack: due to the addition of a fourth beacon, the bacteria move toward another location; in this way, they do not collect the data and the storage system cannot be used.
%
%\begin{figure}[ht]
%\center
%\includegraphics[trim=0mm 0mm 0mm 0mm,clip,width=0.45\textwidth]{figures/spoofing-crop.pdf}
%\caption{Example of Spoofing scenario.}
%\label{fig:spoofing}
%\end{figure}

\subsubsection{Sniffing}\label{subsec:sniffing}
To extract data from the biological storage system, the bacteria need to pick-up the data contained in the clusters and to reach the machine that performs DNA sequencing. During the trip from the archives to the sequencer, an attacker could place a swarm of bacteria in the middle. This would cause the bacteria to conjugate with the ones carrying the encoded information. Consequently, the malicious user can obtain the encoded information without accessing the machine connected to the DNA sequencer. In a network scenario, it would be the equivalent of listening on a channel and sniffing packets.

\section{DoS information metrics}\label{sec:sec_metr}
One way to evaluate an experiment is to define a \textbf{metric}, a standard for measuring a particular characteristic of the experiment. For example, in the field of cybersecurity it is important to measure the damage that an attack (e.g., a DoS) can produce to an infrastructure, or the robustness of the infrastructure in general. Each metric is strictly correlated to what it measures. Therefore, the same metric can perform differently based on the scenario in which it is applied.
%The most commonly used detection techniques against Denial of Service (DoS) attacks are based on
%
%\begin{itemize}
%\item \textit{thresholds:} whenever the incoming traffic overcomes a fixed threshold, we raise a warning;
%\item \textit{signature:} if most of the overloading traffic is coming from a single user, it is very likely that the user may be the attacker, so we raise a warning;
%\item \textit{attributes:} different features of the traffic make us doubt about its intention (e.g., a huge flow of packets of the same size).
%\end{itemize}

It is possible for an attacker to elude detection metrics. For example, if we perform a DoS attack and we do not want to be tracked, we can use multiple machines to conduct the attack. By distributing the outgoing traffic over different IPs we make it harder for the victim to trace us, achieving a Distributed DoS (DDoS). Another way of eluding DoS detection metrics is to maintain a low-rate of traffic sent by the malicious computers, so that it negatively affects the service without exceeding the warning threshold.

Therefore, we need a tool that can overcome similar scenarios.
Xiang et al.~\cite{Xiang} developed an algorithm based on statistical methods to detect a low-rate DDoS attack and traceback the IP of the attacker. Given a Local Area Network (LAN) and a supervisor that monitors the network traffic, their algorithm is based on the following assumptions:

\begin{enumerate}
\item the supervisor has full control of all the routers;
\item they extracted an effective feature (e.g., IP addresses) of network traffic to sample its probability;
\item the supervisor obtained and stored the average traffic of the normal, as well as the local thresholds $\sigma_{f_i}$ and $\sigma_{f(R_i)}$ on its own routers in advance;
\item on all routers, the traffic follows Poisson distribution and the normal traffic follows a Gaussian noise distribution. As stated by Xiang et al.~\cite{Xiang}, it is widely accepted that the Poisson distribution function can simulate the DDoS attack traffic in aggregation and the fractional Gaussian noise function can be simulate real network traffic in aggregation.
\end{enumerate}

The authors redefine two metrics commonly used in information theory: \textit{generalized entropy} and \textit{information distance}. \textbf{Generalized entropy} is a measure of uncertainty associated with a random variable. The more random the information, the bigger the entropy; viceversa, the greater the certainty related to the information, the smaller the entropy. Given a set of events $\{ x_1, x_2, x_3, \ldots, x_n\}$, their associated probabilities ${p_1, p_2, p_3, \ldots, p_n}$ and the following property:

\begin{equation}\label{eq:prob_prop}
0 \leq p_i \leq ~ \forall i = 1 \ldots n, \ \sum_{i=1}^{n} p_{i} = 1,
\end{equation}

the generalized entropy is defined as follows:

\begin{equation}\label{eq:gen_entr}
H_{\alpha}(x) = \frac{1}{1 - \alpha} log_2 \left(\sum_{i=1}^{n} p_{i}^{\alpha} \right), \alpha \neq 0,
\end{equation}

where $\alpha$ is the order of the entropy. When $\alpha \rightarrow 1$, the formula converges to the Shannon entropy:

\begin{equation}
H_1(x) = - \sum_{i=1}^{n} p_i log_2 p_i.
\end{equation}

One of the most important properties of the generalized entropy is that, given $\alpha > 1$, it increases the deviation between the different probability distributions, compared to the Shannon entropy~\cite{Prob1}~\cite{Prob2}. A high probability event contributes more to the final entropy than to the Shannon entropy when $\alpha > 1$, and a low probability event contributes more when $\alpha < 1$. Consequently, we can obtain different entropy values based on the different values of $\alpha$.

Finally, the \textbf{information distance} measures the \textit{divergence} between two probability distributions. Let $P = (p_1, p_2, \ldots, p_n)$ and $Q = (q_1, q_2, \ldots, q_n)$ be two discrete complete probability distribution with the same properties described in Equation~\ref{eq:prob_prop}. The information distance can be calculated as:
\begin{equation}
D_{\alpha}(P || Q) = \frac{1}{\alpha - 1} log_2 \left( \sum_{i=1}^{n} p_{i}^{\alpha} q_{i}^{1 - \alpha}  \right), \alpha \geq 0.
\end{equation}

Note that $D_{\alpha}(P || Q) \neq D_{\alpha}(Q || P)$. In other terms, the information distance is not symmetric. Based on the value of $\alpha$, one of the two distributions (i.e., the one to the power $(1 - \alpha)$) may not be able to contain events where the associated probability is equal to 0. The authors define such distribution as \textit{continuous}. For $\alpha \rightarrow 1$, the information distance becomes the Kullback-Leibler divergence:

\begin{equation}
D_1(P || Q) = \sum_{i=1}^n p_i log_2 \left(\frac{p_i}{q_i}\right).
\end{equation}

In~\cite{Xiang}, Xiang and colleagues modified the information distance equation to satisfy a few properties (namely, additivity, asymmetry and incresing function of $\alpha$) in order to make it compliant to the formal definition of metric. The final result is given in Equation~\ref{eq:inf_dist_metr}:

\begin{multline}\label{eq:inf_dist_metr}
D_{\alpha}(P, Q) = D_{\alpha}(P || Q) + D_{\alpha}(Q || P) = \\
\frac{1}{\alpha - 1} log_2 \left( \sum_{i=1}^{n} p_{i}^{\alpha} q_{i}^{1 - \alpha} \cdot \sum_{i=1}^{n} q_{i}^{\alpha} p_{i}^{1 - \alpha}  \right).
\end{multline}

In this case, none of the two probability distributions can contain events with an associated probability equals to zero. In the end, Xiang et al. used the information distance metric to develop a collaborative DDoS attack detection algorithm, which can be found in~\cite{Xiang}.

%In the following sections, we see how a malicious user can deploy an attack on a DNA archival device and how the two metrics defined in Equations~\ref{eq:gen_entr} and ~\ref{eq:inf_dist_metr} can help us to detect this scenario.

\subsection{Metrics as detection mean}\label{sec:metr}

%Let's recall the two different metrics defined by Xiang et al.~\cite{Xiang} that we described in Section~\ref{sec:sec_metr}: generalized entropy and information distance. From Equation~\ref{eq:gen_entr}, the generalized entropy is defined as
%
%\begin{equation*}
%H_{\alpha}(x) = \frac{1}{1 - \alpha} log_2 \left(\sum_{i=1}^{n} p_{i}^{\alpha} \right),
%\end{equation*}
%
%
%while the latter - i.e., information distance in Equation~\ref{eq:inf_dist_metr} - is
%
%\begin{multline*}
%D_{\alpha}(P, Q) = D_{\alpha}(P || Q) + D_{\alpha}(Q || P) \\
%= \frac{1}{\alpha - 1} log_2 \left( \sum_{i=1}^{n} p_{i}^{\alpha} q_{i}^{1 - \alpha} \cdot \sum_{i=1}^{n} q_{i}^{\alpha} p_{i}^{1 - \alpha}  \right).
%\end{multline*}

Generalized entropy and information distance are based on probability distributions of values, while in our case we have a distribution of value over time. However, a distribution over time can be converted to an approximation of a probability distribution. In fact, given the initial number of bacteria $N_i$ and the total simulation time $T$, we can calculate the probability of observing a certain number of bacteria $b$ as:
\begin{equation}
P(x = b) = \frac{\sum_{t=0}^{T} B(t) = b}{N_i},
\end{equation}

where $B(t)$ is the number of bacteria counted at time $t$ and:

\begin{equation}
(B(t) = b) = \begin{cases}
1 & \mbox{if there are \textit{b} bacteria at time t,}\\
0 & \mbox{otherwise.}
\end{cases}
\end{equation}

The main idea is to use the two metrics to detect if an attack on our archive is taking place. Intuitively, the traffic (i.e., the number of bacteria and how they are distributed over time) is different whether the system is functioning normally or it is under attack. As a result, we define the Algorithm~\ref{alg:distrib} to convert the time sequence into a (sample of) probability distribution. The reason behind this pre-processing is that we wait for bacteria to swim towards their destination, which can take a long time. By defining a long period of time $T$, we can assume a stable mean of bacteria that reach their destination. In this way, we can use the generalized entropy and information distance in our scenario.

\begin{algorithm}[htb]
%\SetAlgoNoLine
\KwIn{$N_i$ number of bacteria, $D$ array of distribution over time.}
\KwOut{$P$ array of probabilities.}
$v \leftarrow 0$\;
\While{$v \leq N_i$}{
		$P(v) \gets countValueInArray(v, D)$\;
		$v \gets v + 1$\;
}
$P \gets divideElementWise(P, length(D))$\;
\caption{Converting a time series to an approximation of a probability distribution.}
\label{alg:distrib}
\end{algorithm}

As long as the sum of these probabilities is not zero, we can calculate the entropy and measure its changes due to an intrusion in the archive. Based on the definition, generalized entropy gives more relevance to rare events: the less likely an event, the bigger the entropy. This implies that when we have a really low number (e.g., 10) of outgoing ``packets'' (i.e., bacteria reaching the destination area) and we lose one of them, the event becomes very frequent ($1/10 = 10\%$) and the generalized entropy is less relevant compared to a scenario where we have a loss of 1 packet over 150 ($1/150 = 0.\overline{66}\%$).

Nevertheless, we need to make some changes in order to use the generalized distance. One of the main assumption made by Xiang et al.~\cite{Xiang} is that the probability distributions involved in the calculation must be continuous. This is not our case, because it is very likely that two different data extractions lead to two different time series of bacteria, due to the randomness in bacteria movement. As a result, it may happens that a certain value never appears in one series, while it does in other series. 

There are two possible way of adapting the metric to our case: using dummy values and transforming the distributions to make them continuous. The usage of dummy values implies substituting each 0 that make the distribution non-continuous with a fixed value (e.g., $10^{-3}$). However, the insertion of a low number would change significantly the value of the metric; on the contrary, a high number would invalidate the fact that the sum of all probabilities is equal to 1. Thus, truncating the distributions in order to make them continuous is a better option. Let us suppose that our distribution is composed of 4 values and their associated probabilities $\{(v_0, p_0), (v_1, p_1), (v_2, p_2), (v_3, p_3)\}$. While it is true that the removal of an event from the distribution (e.g., $(v_3, p_3)$) implies a loss of information, such a loss would not be so impactful. Indeed, the information that a specific event contains is also partially contained in other events: for example, we know that $p_3 = 1 - \sum_{i=0}^{2} p_i$. Algorithm~\ref{alg:make_cont} describes how, given two probability distributions, we can make them continuous according to Xiang et al.'s~\cite{Xiang} definition.

\begin{algorithm}[htb]
%\SetAlgoNoLine
\KwIn{$P$ and $Q$, arrays of probabilities.}
\KwOut{$Pc$ and $Qc$, arrays of continuous probabilities.}
$Pc, Qc \leftarrow emptyArray()$\;
$len \gets min(length(P), length(Q))$\;
$i \gets 0$\;
\While{$i < len$}{
		\If{$P(i) \neq 0$ \textbf{and} $Q(i) \neq 0$}{
			$Pc \gets append(Pc, P(i))$\;
			$Qc \gets append(Qc, Q(i))$\;
		}
		$i \gets i + 1$\;
}
\caption{Converting two probability distributions into two continuous probability distributions.}
\label{alg:make_cont}
\end{algorithm}

In this way, we can use the distance metric to measure the divergence by comparing two different scenarios. This implies that the distance metric is a better way of measuring the diversity of two events, because it compares them directly, while the generalized entropy compares their respective values.

\section{Machine learning for detection}\label{sec:ml}
Despite the appropriateness of a metric, in most cases its mathematical development requires time and resources from researchers. An automated approach would be better in terms of efficiency and efficacy, given that it could extract abnormalities from the data and automatically adapt to new scenarios.

In the cybersecurity field, one of the main tools recently used for detection algorithms is \textbf{machine learning} (ML)~\cite{bookml1}~\cite{bookml2}. For example, ML is used for intrusion detection~\cite{ML1}, as well for DoS~\cite{ML4} and DDoS detection~\cite{ML2}~\cite{ML3}. By using a learning approach, we do not need to develop a different metric for each kind of attack. Instead, we can give a lot of examples (i.e., behaviour of the system under normal circumstances and under attack) to a machine learning algorithm, allowing it to learn to distinguish different scenarios. Moreover, if the algorithm has been trained enough, we could potentially reuse it for different kind of attacks.

Machine learning is usually divided in two main categories: supervised and unsupervised learning. In supervised learning, we have a set composed of examples tuples $\{(x_0, y_0), (x_1, y_1), \cdots (x_n, y_n)\}$ where $x_i$ is the \textit{feature} of the example, and $y_i$ is its label (i.e., an assigned category). Usually, the task of a supervised learning algorithm is to minimize the error related to a function that involves these examples, in order to be able to predict categories (classification) or values (regression). On the other hand, unsupervised learning does not require any category, and the categorization is based upon similarities among the features. %Ideally, the best scenario would be the case in which we do not have to assign categories to the different examples and the algorithm learns to distinguish them without external support.
Ideally, unsupervised learning algorithms learn to categorize the example without any external support.
However, in some cases it is really hard to distinguish similar examples without any indications, because features can be really similar even when they indicate different classes. Moreover, given the pace of virtual simulations, we can generate samples to train the algorithm without the necessity of observing a real system being attacked.

In our case, we can obtain two different information from our simulations: the distribution of bacteria over time, and the probability of observing a specific amount of bacteria over the whole simulation. In addition, we also know the number of bacteria used to retrieve the data and the number of malicious bacteria. 
Consequently, we define three different types of features that we will feed to our algorithm: 
\begin{enumerate*}[label=(\roman*)]
\item the number of bacteria that were able to reach their destination in an interval between two sampling periods,
\item the cumulative number of bacteria that reached their destination up to a specific moment, and
\item a sample of probability distribution, as described in Section~\ref{sec:metr}.
\end{enumerate*}
From now on, we will refer to these features with the names of ``count'', ``sum'' and ``sample''.
Using this data, our goal is to train a machine learning algorithms (e.g., Logistic Regression, Support Vector Machine, or Neural Networks) capable of distinguishing benevolent and malicious traffic in our system during a DoS attack.
%We also explore the capability of such algorithms to distinguish malicious traffic based on the number of attackers (e.g., 200 vs 1000 attackers).

In the following sections, we walk through the steps for pre-processing the data, selecting and evaluating the model, and predicting the results.

%Let's recall from Section~\ref{subsec:att_sim} which data we obtain from the simulations. Given the range of good bacteria $(10,150,10)$ and the range of malicious bacteria $(0, 1900, 100)$, we have $15 \cdot 20 = 300$ different scenarios. For each of these, we conduct 10 runs in order to average the results, leading to a total of 3000 simulations.

\subsection{Pre-processing}

One of the most important ML phases is data \textit{pre-processing}. Usually, it is composed of four steps:
\begin{enumerate*}[label=(\alph*)]
\item \textbf{cleaning:} detecting, removing, and correcting corrupt data;
\item \textbf{integration:} merging in a proper way different kind of data, in order to obtain a unique data set;
\item \textbf{transformation:} changing data values to meet some requirements (e.g., we want features with a specific average and standard deviation), or removing noise;
\item \textbf{reduction:} removing from the data set entries that are useless/redundant.
\end{enumerate*}

In our case, we do not need to go through the integration, but we need to address the other three points. Given the different numbers of legitimate bacteria for the simulations, we obtain different probability distributions from Algorithm~\ref{alg:distrib}. For a scenario that has a lower number of bacteria than other cases, we do not calculate the probability for the same range of values. For example, if a scenario has $n_{1}^{max} = 50$ bacteria and the other scenario has $n_{2}^{max} = 100$, the probability distribution for the first one does not include the probabilities for values greater than 50. Consequently, we need to fill in the missing probability values with zeros. In addition, we normalize all the features so that they follow a normal distribution $N(0,1)$. As we described in Algorithm~\ref{alg:make_cont}, we also remove all the columns that contain zero in all the entries, both from the left and the right side of the data set, in the same way trimming a string would remove the whitespaces at the beginning and at the end of it.

\subsection{Model selection}\label{subsec:mod_sel}
Different predictors have different characteristics. Some of them produce poor results if they are fed with small data sets, while others can perform good even with small data sets. Other models are suited for binary classification (i.e., two possible outcomes), while others can distinguish among more classes. Therefore, we need to choose the best predictor for our specific task.

Our main goal is to detect whether our system is under attack or not. Consequently, we have to perform a binary classification where \textit{0} indicates \textit{under attack} and \textit{1} stands for \textit{normal traffic}. We test five different classifiers: \textit{Support Vector Machine}, \textit{Multi-layer Perceptron}, \textit{Random Forest}, \textit{K-Nearest Neighbors}, and \textit{Logistic Regression}.
%However, it can be useful to distinguish the level of the attack in order to deploy different mitigation strategies. Consequently, we need to use a multiclass predictor. We test six different classifiers: Support Vector Machine, Multi-layer Perceptron, Random Forest, K-Nearest Neighbors, Extra Trees and Logit; all of them are included in the Scikit-learn~\cite{scikit-learn} Python library.

\subsection{Metrics}\label{subsec:metrics}
As we previously mentioned, in order to measure the quality of our choices and results, we need to define some metrics. There are some metrics widely used in machine learning which consider the capability of the predictor to produce correct results. In binary classification, given an example $(x_i, y_i)$ and its predicted value $y'_{i}$, there are four possible scenarios:
\begin{enumerate*}[label=(\roman*)]
\item \textbf{T}rue \textbf{P}ositive (TP), where $y_i = y'_i = 1$;
\item \textbf{T}rue \textbf{N}egative (TN), $y_i = y'_i = 0$;
\item \textbf{F}alse \textbf{P}ositive (FP), $y_i = 0$ and $y'_i = 1$;
\item \textbf{F}alse \textbf{N}egative (FN), $y_i = 1$ and $y'_i = 0$.
\end{enumerate*}

From these four possible outcomes, we can define the following metrics:
\begin{itemize}
\item \textbf{Accuracy ($\alpha$):} $\frac{TP+TN}{TP+FP+TN+FN}$;
\item \textbf{Precision ($\pi$):} $\frac{TP}{TP+FP}$;
\item \textbf{Recall ($\rho$, also called Sensitivity or True Positive Rate):} $\frac{TP}{TP+FN}$;
\item \textbf{F1 score:} $2 \frac{\pi \rho}{\pi + \rho}$;
\item \textbf{Specificity (or True Negative Rate):} $\frac{TN}{TN+FP}$;
\item \textbf{False Positive rate:} $1 - Specificity$;
\item \textbf{AUC (Area Under Curve):} the area under a curve that represents the ratio between correct and wrong predictions (e.g., precision-recall curve).
\end{itemize}

Accuracy is a general score describing how many guesses from the predictor are correct, but it is not reliable in case of unbalanced data sets. Sensitivity (i.e., recall) and specificity indicate, respectively, the proportion of positives and negatives that are correctly identified as such. Precision represents the number of actual positives, among all the predicted positives. The F1 score is a combination of precision and recall, which also describe the accuracy of the predictor in case of an unbalanced dataset.
Finally, the Receiver Operating Characteristic (ROC) curve plots True Positive Rate (TPR) versus False Positive Rate (FPR) at different discrimination thresholds - i.e., a value used to determine whether to classify a sample as one class or another. Thus, the ROC curve represents the TPR as a function of the FPR. Similarly, the Precision-Recall curve plots precision versus recall while varying the discrimination threshold - this particular curve is ideal for unbalanced datasets. By calculatitng the Area Under Curve (AUC) for these two different measures, we obtain an aggregate measure of performance across all possible classification thresholds. We define the AUC for ROC as AUROC and the AUC for Precision-Recall curve as AUPRC.
%For the multiclass predictors, Scikit-learn~\cite{scikit-learn} defines some variants of the previous metrics in order to combine the scores for different classes, adding the following affixes: \textit{micro}, \textit{macro}, \textit{samples} and \textit{weighted}. In particular, the \textit{weighted} variant calculates the metrics for each label and find their average weighted by the number of occurrences of the label in the true data.

\subsection{Training and parameter tuning}
Usually, before training a machine learning algorithm we must divide the data set in three different subsets:
\begin{itemize}
\item \textbf{training set:} the part of the data used to train the algorithm;
\item \textbf{validation set:} the subset used to choose the model, and tune its hyperparameters;
\item \textbf{test set:} the data used to evaluate the efficacy of the algorithm, after it has been trained.
\end{itemize}

In this way, we evaluate the algorithm with neutral data that was not part of its training. Moreover, the division of training and test set reduce the risk of \textit{overfitting}, which is the phenomenon of producing an analysing that corresponds too closely to a specific data set. One further step for avoiding overfitting is to use \textit{cross validation}. In particular, \textit{k-fold cross validation} is a technique that splits the data into $K$ chunks of the same size, trains the model with $K-1$ chunks, and uses the remaining data as validation set. These steps are then repeated for each chunk. Cross validation helps both with overfitting and parameter tuning, increasing the flexibility of an evaluation. 

In our case, we split our data sets using the 70\% of it as training (and validation) set, and the remaining 30\% as test set. We implement this division using the \texttt{train\_test\_split} function from Scikit-learn machine learning library~\cite{scikit-learn}. In addition, we use the \texttt{GridSearchCV} class to execute an exhaustive parameters search using cross validation. Table~\ref{tab:param_tun} lists all the parameters that we investigate using the cross validation method.

\begin{table}[htb]
\centering
\caption{Parameters inspected during cross validation. The parameters name follows the nomenclature used by Scikit-learn~\cite{scikit-learn}}
\begin{tabular}{ll}
\hline
\textbf{Model} & \textbf{Parameters}                          \\ \hline
MLP            & hidden\_layer\_sizes, solver, learning\_rate \\
KNN            & n\_neighbors, weights, algorithm             \\
SVM            & C, gamma, kernel, max\_iter                  \\
%ETC            & n\_estimators, criterion, max\_features     \\
RF             & n\_estimators, max\_features                 \\
LR             & fit\_intercept                               \\ \hline
\end{tabular}
\label{tab:param_tun}
\end{table}

\section{Evaluations and Results}\label{sec:sim}
We defined different tools (i.e., metrics and machine learning algorithms) to detect if a malicious user is trying to compromise a DNA-based archive. In this section, to assess the quality of such tools, we run a number of simulations that replicate the scenarios illustrated in Section~\ref{sec:vuln}. Each simulation represents one possible attack, demonstrating how this vulnerability can affect the storage system. In the end, we implement the different detection techniques described in Section~\ref{sec:sec_metr} and Section~\ref{sec:ml} to verify their efficacy.

For each of the following simulations, the maximum amount of bacteria contained in each cluster is 50, as mentioned by Tavella et al.~\cite{Tavella}. We conform to this parameter in order to enable comparison between the data obtained during an undisturbed run of the system and the following simulations. If the bacteria are not able to retrieve the whole file within 120 virtual minutes (i.e., two hours in the simulation), we stop the simulation. Finally, we indicate the intervals as tuples $[l, u, s]$ where $l$ is the lower bound, $u$ is the upper bound (both included in the interval), and $s$ is the step size. For example, the tuple $[10, 30, 10]$ indicates a range composed of the values $10$, $20$, and $30$.

\subsection{Attacks simulations}\label{subsec:att_sim}
In this section, we describe the simulations we conducted for evaluating the disruption of a DoS attack. In addition, we define the dangerousness of an attack in terms of delay and information loss, which is the incapability of retrieving data within a fixed amount of time.

We have previously illustrated in Figure~\ref{fig:dos} how, during a Denial of Service attack, malicious bacteria move towards the clusters and conjugate with the bacteria containing the encoded data, obstructing the legitimate bacteria. In our simulations, we decided to vary the number of legitimate bacteria in the range $[10, 150, 10]$, according to the simulations conducted in~\cite{Tavella}, and the number of malicious bacteria in the range $[0, 1900, 100]$. Until the amount of legitimate and malicious bacteria is lower than the total amount of bacteria in the cluster, it is not possible to detect any kind of attack because the cluster is not running at maximum conjugation rate. However, since the conjugation has a probability approximated by a Normal distribution with mean 0 and standard deviation 1, it is not certain that we are able to detect any difference even if the sum of the two types of bacteria is slightly greater than the threshold.

\begin{figure}[htb]
\centering
\includegraphics[trim=0mm 0mm 0mm 0mm,clip,width=0.45\textwidth]{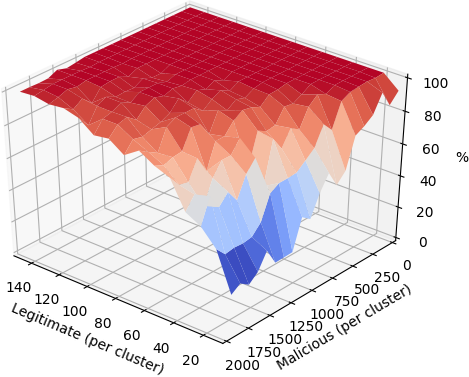}
\caption[Correlation among the number of engineered motile bacteria, number of attackers and percentage of file retrieved.]{Correlation among the number of engineered motile bacteria, number of attackers and percentage of file retrieved. If the percentage is lower than 100\%, it means that the engineered motile bacteria were not able to retrieve the whole file in less than 120 minutes.}
\label{fig:perc_3d}
\end{figure}

\begin{figure}[htb]
\centering
\includegraphics[trim=0mm 0mm 0mm 0mm,clip,width=0.45\textwidth]{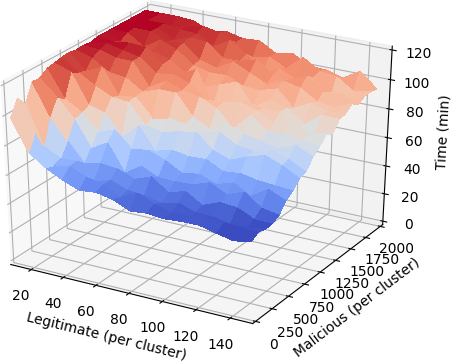}
\caption[Correlation among the number of engineered motile bacteria, number of attackers and time necessary to retrieve the whole file.]{Correlation among the number of engineered motile bacteria, number of attackers and time necessary to retrieve the whole file. If the time is greater or equal to 120 minutes, then the engineered motile bacteria were not able to retrieve all the information.}
\label{fig:time_3d}
\end{figure}

Figure~\ref{fig:perc_3d} shows how the percentage of retrieved file changes with respect to the number of legitimate and malicious bacteria. Even with a really high amount of attackers, it is sufficient to use 150 retrievers to extract the whole file from the archive. This shows that the percentage of retrieved file is not a good measure to detect an attack. Figure~\ref{fig:time_3d} illustrates how the time to retrieve the file is affected by the number of bacteria. In this case, we can see how the number of attackers affects the final time, increasing it from ~40 minutes (150 retrievers vs. 0 attackers) to ~100 minutes (150 retrievers vs. 1900 attackers). In many cases, such as with attackers $> 1000$ and retrievers $< 100$, the average time to retrieve the file gets really close to the 120 minutes threshold. 

In Section~\ref{subsec:sniffing}, we hypothesised that a malicious user could steal a copy of the archived information, without being noticed. From the results in Figure~\ref{fig:perc_3d} and Figure~\ref{fig:time_3d}, we can deduce that an attacker could introduce a small number of bacteria and obtain the archived data, while keeping the delay low enough to avoid being detected.

\subsection{Evaluation of information metrics}
Before analysing the importance of the metric order $\alpha$, let us recall its role. Given a metric, such as the generalized entropy, its order $\alpha$ increases/reduce the importance of an event. Let $p$ be the probability of an event, with $0 \leq p \leq 1$. If $p \neq 1$, there are two different scenarios:
\begin{enumerate}
\item if $\alpha > 1$, then $p > p^{\alpha}$;
\item otherwise ($\alpha < 1$), $p < p^{\alpha}$.
\end{enumerate}

Consequently, whenever we impose the metric order below 1, we increase the value corresponding to the probability of an event. With really small values of $\alpha$, the closer the probability is to 0, the bigger is the amplification. For example with $\alpha = 0.1$ and $p = 0.1$, $0.1^{0.1} - 0.1 \approx 0.69$, while with $\alpha = 0.1$ and $p = 0.5$, $0.5^{0.1} - 0.5 \approx 0.43$. In light of this, if we use an order which is too small, we risk to flatten the diversity between probabilities.

\begin{figure}[htb]
\centering
\includegraphics[trim=0mm 0mm 0mm 0mm,clip,width=0.45\textwidth]{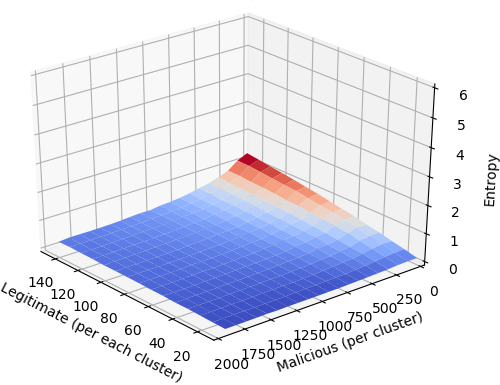}
\caption{Generalized entropy as a function of benevolent and malicious bacteria. The sampling period is equal to 10 seconds and $\alpha = 2$.}
\label{fig:alpha2e}
\end{figure}

Here we present the most significant results; for the detailed results, we refer the reader to Appendix~\ref{app:met}.
We test four different orders: $0.5, 2, 5, 10$.
We notice that there is no significant difference in using $\alpha = 2$, $\alpha = 5$ or $\alpha = 10$, so we decide to remove 5 and 10 (i.e., the values that could cause unwanted spikes in our curves)from the possible orders for the entropy. Figure~\ref{fig:alpha2e} shows how the metric changes with different numbers of legitimate and malicious bacteria, for $\alpha = 2$.

\begin{figure}[htb]
\centering
\includegraphics[trim=0mm 0mm 0mm 0mm,clip,width=0.45\textwidth]{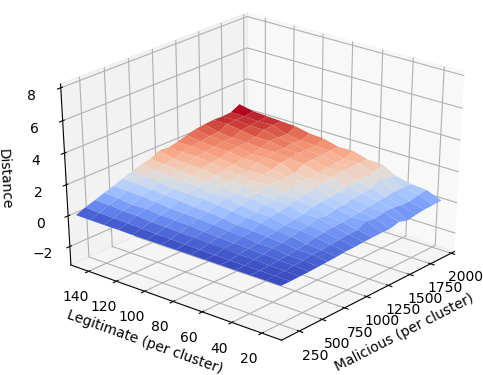}
\caption{Information distance as a function of benevolent and malicious bacteria. The sampling period is equal to 10 seconds and $\alpha = 2$.}
\label{fig:alpha2d}
\end{figure}

Through our experiments we also found out that $\alpha$ severely affects the information distance. When the order is below 1, it drastically flattens the differences. When the order is too high (e.g., 10) the information distance lacks of monotonous behaviour, presenting a lot of spikes over the curve. As a consequence, we discard 0.5 and 10 as values for the information distance. Figure~\ref{fig:alpha2d} illustrates the behaviour of the metric with order $\alpha = 2$.

To summarize, we have two order values for the entropy ($0.5$ and $2$) and for the information distance ($2$ and $5$) which present different values for the respective metrics based on whether the system is under attack or not. 
%However, these two measure express really different results: generalized entropy is affected more by the number of legitimate bacteria than the number of malicious, as shown in Figure~\ref{fig:alpha2e}; on the other hand, variations in the information distance value are related exclusively to the number of attackers - Figure~\ref{fig:alpha2d}.

\begin{figure}[htb]
\centering
\includegraphics[trim=0mm 0mm 0mm 0mm,clip,width=0.45\textwidth]{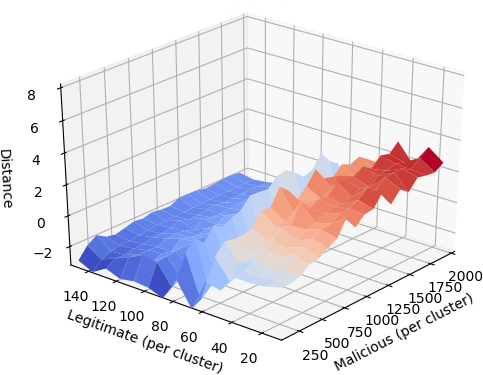}
\caption{Impact of number of legitimate/malicious bacteria on information distance with $\alpha = 2$ and sampling period of 240 seconds. Compared to Figure~\ref{fig:alpha2d}, the sampling period has a significant impact.}
\label{fig:freq6d1}
\end{figure}

Finally, our sampling frequency plays a major role in the values of our metrics: a low sampling frequency can hide the differences between distributions, while a high sampling frequency is hard to implement from a technical point of view, due to separation of bacteria and count. Therefore, we decide to explore different scenarios where we gradually decrease the sampling frequency. Given 6 different sampling periods - 10, 20, 30, 60, 120, and 240 seconds - we observe in Figure~\ref{fig:alpha2e} no significant variations for the generalized entropy. On the other hand, the information distance is heavily affected by the sampling period, changing its behaviour from detecting changes in the number of malicious bacteria while using a small sampling period (Figure~\ref{fig:alpha2d}) to detecting changes in the number of legitimate ones while using a larger sampling period (Figure~\ref{fig:freq6d1}).

\subsection{Evaluation of Machine Learning}
Here we describe in detail the results for binary classification. For each metric, we discuss which are the best features and models to use in order to obtain the best classificator.

Let us recall that the accuracy is defined as the ratio of correct predictions over the whole set of predictions, It gives us an idea about the efficacy of our predictor. However, if one class cardinality is much larger than the other, correct predictions over the majority class have a stronger influence over the accuracy. Thus, we need to consider metrics that separates correct predictions for positive and negative samples, such as sensitivity, specificity, and precision. Hence, we decide to use the AUROC and the Area Under Precision-Recall curve as a combination of the best metrics for our scenario.

\begin{figure}[htb]
\centering
\subfloat[Area Under ROC curve.]{\includegraphics[width=0.45\textwidth]{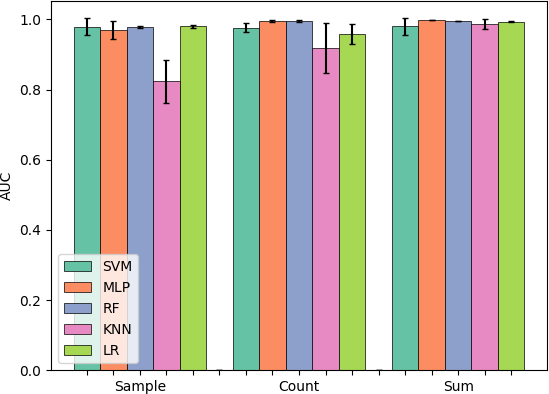}
\label{fig:roc_auc_avg}}
\hfill
\subfloat[Area Under Precision-Recall curve.]{\includegraphics[width=0.45\textwidth]{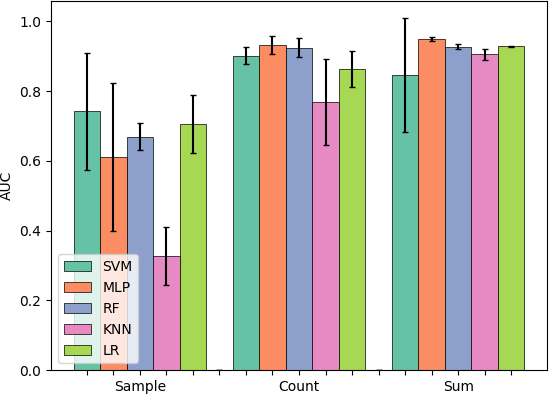}
\label{fig:prec_rec_avg}}
\caption{Mean and standard deviation of the Area Under ROC curve and Precision-Recall curve for different algorithms, calculated using a sampling period of 10, 20, 30, 60, 120 and 240 seconds.}
\label{fig:auc_avg}
\end{figure}

Figure~\ref{fig:auc_avg} shows how the correctness of our predictor is affected by the different features and different sampling periods, using the Area under the ROC curve and Precision-Recall curve. If we take into consideration only the AUROC (Figure~\ref{fig:roc_auc_avg}), we can observe that varying sampling periods does not drastically affect the score - i.e., low standard deviation. The ``sum" feature appears to be the best for all the classifiers. Regarding the different classification algorithms, K-NN is the one with the poorest performance and the highest standard deviation. On the contrary, Random Forest is one of the best algorithms across all the classifiers, with a negligible variance.
We can see that there is no feature that is capable of reducing the standard deviation for all the algorithms simultaneously and K-NN remains the algorithm that performs worse in most of the scenarios. 
Even if we consider a different score, the feature that brings most of the algorithms to the same level is ``sum". In fact, the importance of using this feature is remarked by \Crefrange{fig:over_prec_rec1}{fig:over_prec_rec3} in Appendix~\ref{app:ml}. 
%In particular, Figure~\ref{fig:over_prec_rec3} and Figure~\ref{fig:over_prec_rec1} show that using the ``sum" feature instead of the ``sample" one, the performance of different algorithms can increase from ${\sim} 5$\%, in the case of Support Vector Machine, to ${\sim} 165$\%, using KNN.

Finally, given a sampling period of $T = 10 (s)$, \Crefrange{fig:over_roc_auc1}{fig:over_roc_auc3} and \Crefrange{fig:over_prec_rec1}{fig:over_prec_rec3} describe how every algorithm performs over different features, using ROC curves and Precision-Recall curves. In \Crefrange{fig:over_roc_auc1}{fig:over_roc_auc3} we notice that no algorithm, except for K-Nearest Neighbors, scores an Area Under ROC curve below 0.95. On the other hand, all the algorithms score above 0.98 with the ``sum" feature.

\section{Conclusion}\label{sec:disc}
Technology incredibly evolved and revolutionized society over the last decade. Sometimes technological breakthroughs trigger the necessity for an update of existing tools. In the case of Big Data, one of the obvious consequences is the necessity for new storage devices, able to keep up with the data trend.

With its high capacity/volume ratio, DNA is among the most interesting candidates for solving the storage issue. However, biological devices storing information in their DNA can suffer of security issues ascribed to traditional electronic architectures. %Considering DNA archives based on bacteria (e.g., DN3A~\cite{Tavella}), the new vectors for attacks are biological tools. 
A malicious user can use malicious bacteria to replicate conventional attacks, such as DoS attacks, on a bacterial nanonetworks. In order to mitigate these risks, we need to deploy some countermeasures to prevent and detect these new kind of threats. In this paper, we focus on two different detection techniques: metrics and machine learning. In particular, we adapt to our scenario metrics for detecting traditional DoS attacks.

Applying some changes to the metrics defined by Xiang et al~\cite{Xiang}, we manage to distinguish legitimate traffic from malicious traffic. Moreover, we use machine learning algorithms to perform binary classification using three different features, in order to analyse the bacterial network traffic. Considering the ``sum'' feature, we scored an AUROC over 0.99, and an AUPRC over 0.91, proving that we can reliably distinguish whether the system is under attack or not. We also showed that - due to the nature of our data - K-NN is the worst performing classification algorithm, while RF proved to be most consistent, across different features.

%Although we used different features, when we compare K-NN to all the other algorithms, we did not manage to obtain similar results. We assume this behaviour is caused by two different factors: the data set composition and how the algorithm works. Let us recall that the K-nearest neighbours algorithm calculates the label based on a majority vote amongst the k-nearest examples. When evaluating the probability of an example belonging to a specific class, if there is not a neat distinction between the two (or more) subspaces it is more likely that the example is classified with the label corresponding to the majority class. In our scenario, we have far more examples corresponding to malicious bacteria than legitimate. Thus, it is more likely that using K-NN an example of malicious traffic, yielding to a misclassification.

%As future directions, we would like to investigate the efficacy of the same predictors and similar metrics over different kind of attacks and identify new features that work with a wider range of algorithms.

%\section*{Acknowledgment}


\ifCLASSOPTIONcaptionsoff
  \newpage
\fi

\bibliography{references}
\bibliographystyle{ieeetr}	

\newpage
%\appendix
\begin{appendices}
\section{Metrics results}\label{app:met}
\begin{figure}[h]
\centering
\includegraphics[trim=0mm 0mm 0mm 0mm,clip,width=0.45\textwidth]{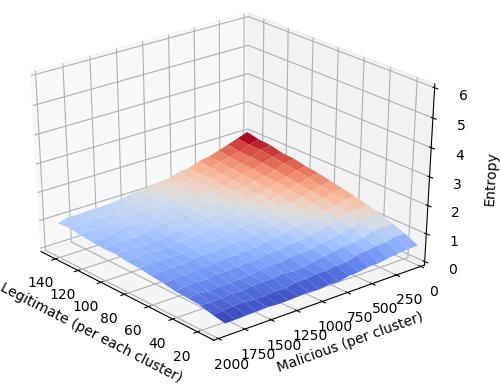}
\caption{Impact of metric order on generalized entropy with sampling period $T = 10 s$ and $\alpha = 0.5$.}
\label{fig:freq1e1}
\end{figure}

\begin{figure}[h]
\centering
\includegraphics[trim=0mm 0mm 0mm 0mm,clip,width=0.45\textwidth]{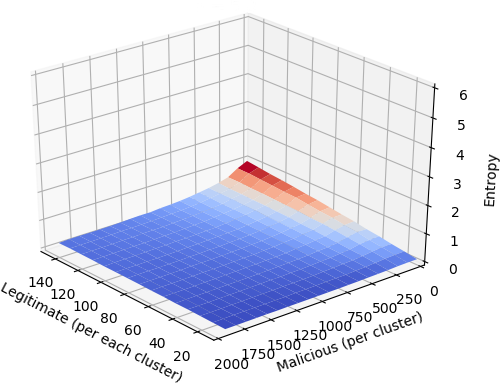}
\caption{Impact of metric order on generalized entropy with sampling period $T = 10 s$ and $\alpha = 5$.}
\label{fig:freq1e3}
\end{figure}

\begin{figure}[h]
\centering
\includegraphics[trim=0mm 0mm 0mm 0mm,clip,width=0.45\textwidth]{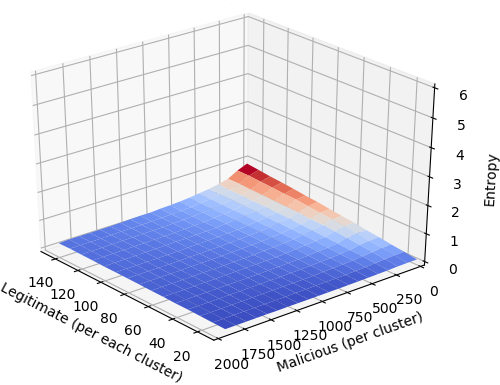}
\caption{Impact of metric order on generalized entropy with sampling period $T = 10 s$ and $\alpha = 10$.}
\label{fig:freq1e4}
\end{figure}

\begin{figure}[!h]
\centering
\includegraphics[trim=0mm 0mm 0mm 0mm,clip,width=0.45\textwidth]{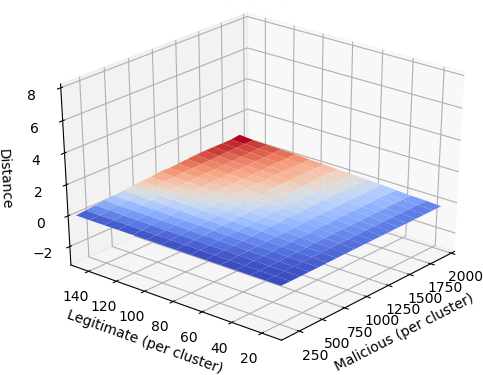}
\caption{Impact of metric order on information distance with sampling period $T = 10 s$ and $\alpha = 0.5$.}
\label{fig:freq1d1}
\end{figure}

\begin{figure}[h]
\centering
\includegraphics[trim=0mm 0mm 0mm 0mm,clip,width=0.45\textwidth]{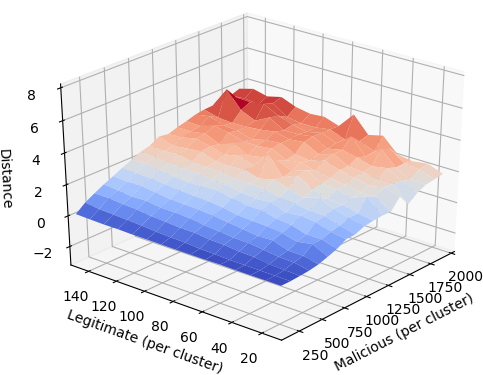}
\caption{Impact of metric order on information distance with sampling period $T = 10 s$ and $\alpha = 5$.}
\label{fig:freq1d3}
\end{figure}

\begin{figure}[h]
\centering
\includegraphics[trim=0mm 0mm 0mm 0mm,clip,width=0.45\textwidth]{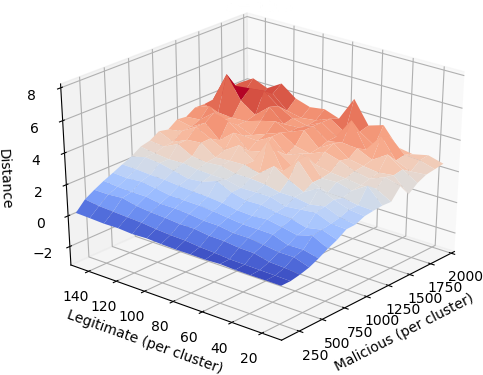}
\caption{Impact of metric order on information distance with sampling period $T = 10 s$ and $\alpha = 10$.}
\label{fig:freq1d4}
\end{figure}

\Crefrange{fig:freq1e1}{fig:freq1e4} and \Crefrange{fig:freq1d1}{fig:freq1d4} show the results we obtained by varying the order $\alpha$ for generalized entropy and information distance. By increasing $\alpha$, the generalized entropy becomes less responsive to variations in the number of malicious bacteria. On the other hand, the information distance lacks of monotic behaviour, as we can see in by the spikes in Figure~\ref{fig:freq1d3} and Figure~\ref{fig:freq1d4}.

\section{Machine learning results}\label{app:ml}
\begin{figure}[h]
\centering
\includegraphics[trim=0mm 0mm 0mm 0mm,clip,width=0.45\textwidth]{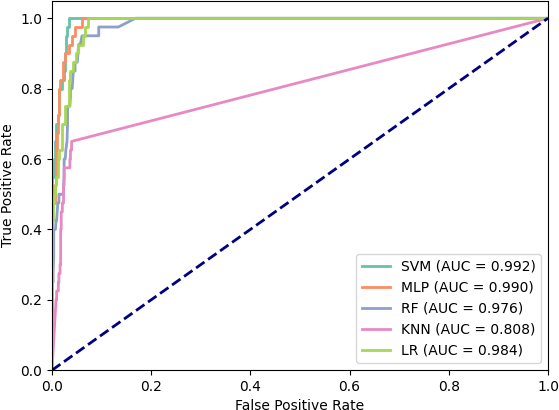}
\caption{ROC curve of different algorithms using the sample feature and $T = 10 s$ as sampling period.}
\label{fig:over_roc_auc1}
\end{figure}

\begin{figure}[h]
\centering
\includegraphics[trim=0mm 0mm 0mm 0mm,clip,width=0.45\textwidth]{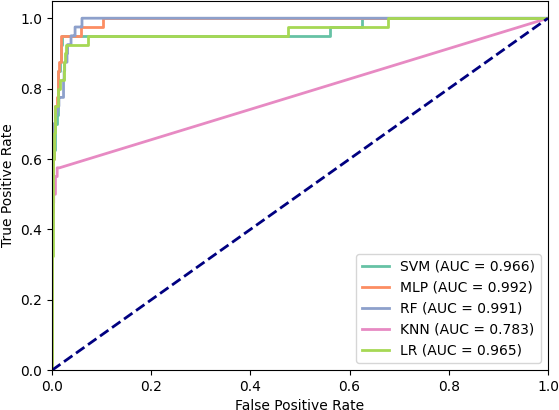}
\caption{ROC curve of different algorithms using the count feature and $T = 10 s$ as sampling period.}
\label{fig:over_roc_auc2}
\end{figure}

\begin{figure}[h]
\centering
\includegraphics[trim=0mm 0mm 0mm 0mm,clip,width=0.45\textwidth]{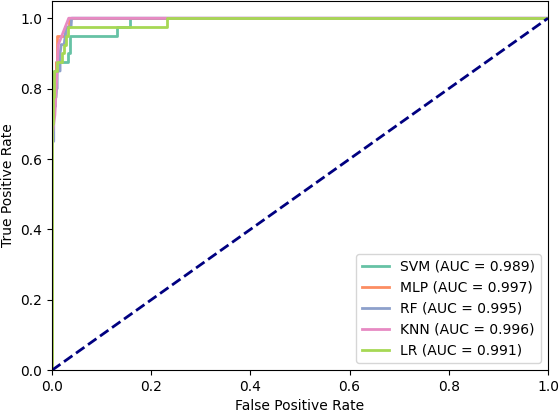}
\caption{ROC curve of different algorithms using the sum feature and $T = 10 s$ as sampling period.}
\label{fig:over_roc_auc3}
\end{figure}

\begin{figure}[h]
\centering
\includegraphics[trim=0mm 0mm 0mm 0mm,clip,width=0.45\textwidth]{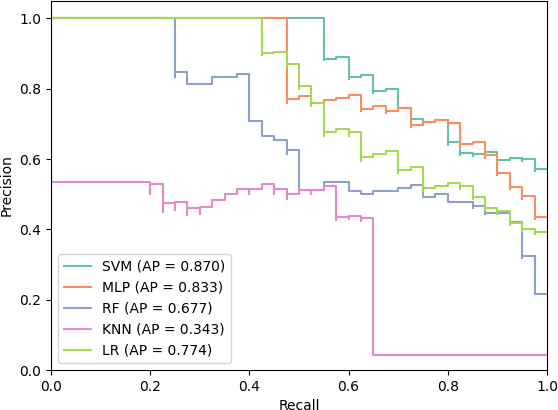}
\caption{Precision-recall curve of different algorithms using the sample feature and $T = 10 s$ as sampling period.}
\label{fig:over_prec_rec1}
\end{figure}

\begin{figure}[h]
\centering
\includegraphics[trim=0mm 0mm 0mm 0mm,clip,width=0.45\textwidth]{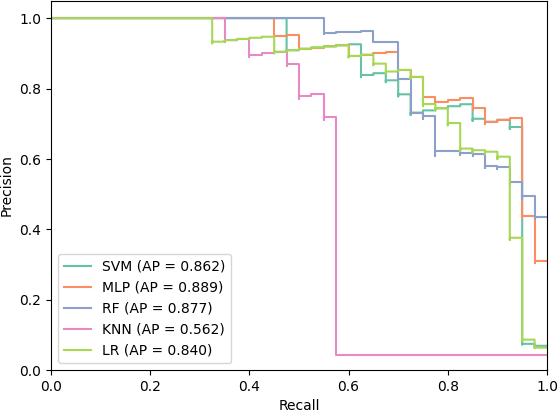}
\caption{Precision-recall curve of different algorithms using the count feature and $T = 10 s$ as sampling period.}
\label{fig:over_prec_rec2}
\end{figure}

\begin{figure}[h]
\centering
\includegraphics[trim=0mm 0mm 0mm 0mm,clip,width=0.45\textwidth]{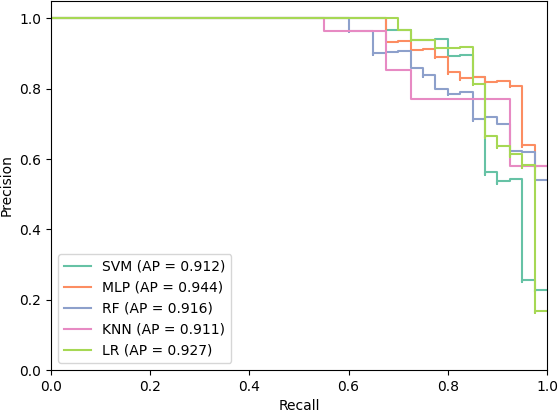}
\caption{Precision-recall curve of different algorithms using the sum feature and $T = 10 s$ as sampling period.}
\label{fig:over_prec_rec3}
\end{figure}

\Crefrange{fig:over_roc_auc1}{fig:over_roc_auc3} and \Crefrange{fig:over_prec_rec1}{fig:over_prec_rec3} illustrate the ROC and Precision-Recall curve for each model, using a sample period of 10 seconds. For each different model and feature, we present the respective score in the legend of the graphs. As previously discussed, we can see that the K-NN algorithm is the one performing worst over different features.

\begin{figure}
\includegraphics[width=0.45\textwidth]{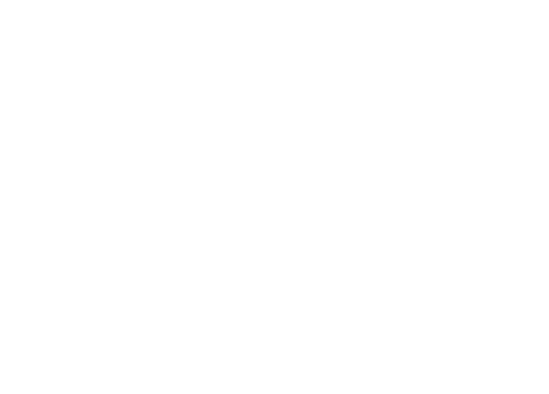}
\end{figure}

\begin{figure}
\includegraphics[width=0.45\textwidth]{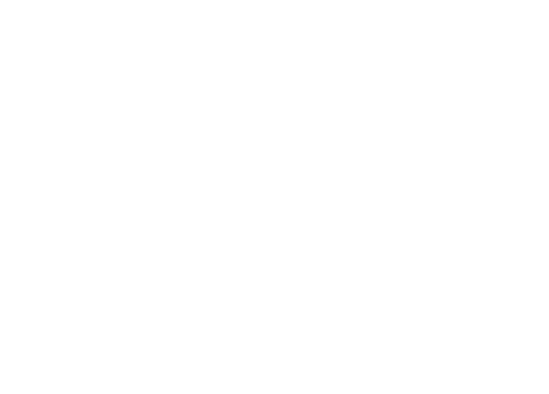}
\end{figure}

\end{appendices}
\end{document}